\documentclass[12pt]{article}
\usepackage{epsfig}
\newcommand{\na}{\mbox{\boldmath $\nabla$}}
\newcommand{\rr}{\mbox{\boldmath $r$}}
\begin{document}
\begin{center}
{\Large\bf Neutrostriction in neutron stars}
\medskip

{\bf Vladimir K. Ignatovich}
\medskip

{\it Frank Laboratory of Neutron Physics of Joint Institute for
Nuclear Research, 141980, Dubna Moscow region, Russia}
\bigskip

\begin{abstract}
It is demonstrated that not only gravity, but also neutrostriction
forces due to optical potential created by coherent elastic
neutron-neutron scattering can hold a neutron star together. The
effect of these forces on mass, radius and structure of the
neutron star is estimated.
\end{abstract}
\end{center}

\section{Introduction}

Interaction of neutrons with matter at low energies is
characterized by optical potential
\begin{equation}\label{1}
V_o(\rr)=\frac{\hbar^2}{2m}4\pi n(\rr)b,
\end{equation}
where $m$ is neutron mass, $n(\rr)$ is atomic density at a point
$\rr$, and $b$ is coherent s-wave neutron-nucleus scattering
amplitude (see, for example~\cite{ig0,ign}). It is important that
though the amplitude $b$ is the result of short range strong
interactions, the potential $V_o$ is the long range one.

Neutron-nucleus scattering amplitude $b$ is of the order of
several fm, the density of matter in the earth conditions is of
the order $10^{23}$ cm$^{-3}$, so the optical potential of matter
is of the order $10^{-7}$ eV.

Interaction of neutrons with neutron matter is also described by
eq. (\ref{1}), and since the amplitude $b$ is negative, the
potential (\ref{1}) of the neutron matter is attractive. The
attractive force $F_o=-\na V_o(\rr)$ is called neutrostriction.

Neutron-neutron s-wave scattering can take place only in singlet
state, and the singlet amplitude $b_s$ at low energies is
$b_s\approx -18$ fm~\cite{huhn}. The coherent amplitude obtained
by averaging over all possible spin states of two neutrons is 4
times lower, therefore $b=-4.5$ fm. For such values of $b$ at star
densities larger than $10^{36}$ neutrons/cm$^{3}$ the potential
$V_o$ is larger than 1 MeV.

Let's compare the total gravitational and optical energies for a
star of radius $R$, mass $M$ and uniform density
\begin{equation}\label{1a}
n=\frac{M}{(4\pi/3)R^3m}.
\end{equation}
The total gravitational energy is
\begin{equation}\label{1b}
U_g=\frac35G\frac{M^2}{R},
\end{equation}
where $G$ is gravitational constant, while the optical energy is
\begin{equation}\label{1c}
U_{opt}=\frac{\hbar^2}{2m}4\pi
nbN=3\frac{\hbar^2}{2m^3}bM^2/R^3,
\end{equation}
where $N$ is the total number of neutrons in the star. For
amplitude $|b|=4.5$ fm we have $U_{opt}>U_g$, when~\cite{conf}
\begin{equation}\label{1d}
R<R_0=\sqrt{\frac{5\hbar^2|b|}{2m^3G}}=\frac{\hbar}{mc}
\sqrt{\frac{5mc^2|b|}{2Gm^2}}\approx 20{\rm\ km},
\end{equation}
which is independent of the star mass $M$. Using the parameter
$R_0$ one can rewrite (\ref{1c}) as
\begin{equation}\label{1cc}
U_{opt}=\frac35GR_0^2\frac{M^2}{R^3}.
\end{equation}

However, the s-wave amplitude $b_s$ is a constant only at low
energies. In neutron stars, where energies are rather high, energy
dependence of $b$ has to be taken into account. This dependence is
well described by the theory of effective radius
\begin{equation}\label{1d2a}
\frac{1}{b_s(E)}=\frac{1}{b_s(0)}\left(1-\frac{1}{2}k^2ab_s(0)\right),
\end{equation}
where $a$ is the effective radius of neutron nucleus interaction,
$a=1.2$ fm, $b_s(0)=-18$ fm, and $k^2=2mE/\hbar^2$. One can now
determine $b$ as the function of energy. This relationship is
shown in (\ref{1d2})
\begin{equation}\label{1d2}
b(E)\equiv\frac{b_s(E)}{4}=\frac{b(0)}{1+Qx^2}.
\end{equation}
In (\ref{1d2}) $x=k/k_c$, $k_c=mc/\hbar=4.8\cdot10^{13}$
cm$^{-1}$, and $Q=a|b_s(0)|k_c^2/2\approx250$. In the degenerate
neutron gas the most important is the energy at the Fermi level:
$E=E_F=\hbar^2k^2_F/2m$, where $k_F$ is the neutron wave number at
the Fermi level. It is related to neutron density by equation
$n=k_F^3/3\pi^2$. Thus $b(E)=b(E_F)$ can be represented as a
function of $n$:
\begin{equation}\label{1d2x}
b(E)\equiv b(n)=\frac{b(0)}{1+Qx^2}=\frac{b(0)}{1+Q(n/n_c)^{2/3}},
\end{equation}
where a unit of density $n_c=k_c^3/3\pi^2\approx3.7\cdot10^{39}$
cm$^{-3}$ is introduced.

In the next section the contribution of the optical potential to
neutron star parameters $R$, $M$ and density distribution $n(r)$
is estimated. The neutron star is considered as a nonrotating
spherical object composed of a degenerate neutron gas at zero
temperature. Calculations were performed with the help of the
Tolman-Oppenheimer-Volkov (TOV) equation~\cite{re}, generalized by
inclusion of neutrostriction forces. We do not take into account
the short range nuclear forces because they come into play only at
nuclear densities $n_N\approx10^{38}$ cm$^{-3}$, while optical
potential is the most important at $n\ll n_N$.

In the third section the optical potential is compared to commonly
used short range nuclear interactions, and in conclusion some
effects are discussed, which can take place in neutron stars, if
energy dependence $b(E)$ contains a resonance.

\section{Neutron star without short range nuclear interactions}

The Tolman-Oppenheimer-Volkov (TOV) equations in nonrelativistic
Newtonian form are
\begin{equation}\label{s1}
\frac{dp(r)}{dr}=-G\frac{\varepsilon(r){\mathcal M}(r)}{c^2r^2},
\end{equation}
\begin{equation}\label{s2}
\frac{d{\mathcal M}(r)}{dr}=\frac{4\pi r^2\varepsilon(r)}{c^2},
\end{equation}
where $\varepsilon(r)$ is energy density of particles given by
their Fermi distribution
\begin{equation}\label{6}
\varepsilon(r)=\frac{8\pi\hbar
c}{(2\pi)^3}\int\limits_0^{k_F(r)}\sqrt{k^2+k_c^2}k^2dk=
mc^2n_{c}\int\limits_0^{x}3\sqrt{u^2+1}u^2du=\varepsilon_{c}f(x),
\end{equation}
$x=k_F/k_c=\sqrt[3]{n/n_c}$, $n_c=k_c^3/3\pi^2$,
$\varepsilon_{c}=mc^2n_{c}=5.6\cdot10^{36}$ erg/cm$^3$, and
$$f(x)=3\int\limits_0^{x}\sqrt{1+u^2}u^2du=$$
\begin{equation}\label{5}
\frac38\left[(2x^2+1)x\sqrt{1+x^2}-\ln(x+
\sqrt{1+x^2})\right]=x^3\cases{1&for $x\to0$\cr3x/4&for
$x\to\infty$\cr}.
\end{equation}

\begin{figure}[t]
{\par\centering\resizebox*{6cm}{!}{\includegraphics{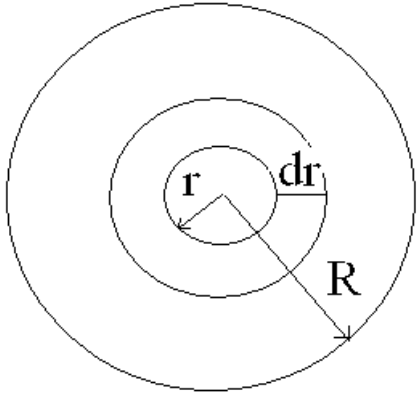}}\par}
\caption{\label{fn1}To derivation of the TOV equation.}
\end{figure}

The first equation (\ref{s1}) represents a condition for a balance
between pressure and gravita\-ti\-o\-nal compression acting on a
mass element $dM=4\pi r^2drmn(r)$ ($m$ is the neutron mass and
$n(r)$ is the particles density) within a thin spherical shell of
thickness $dr$ shown in fig. \ref{fn1}. The force $F_p=4\pi
r^2[p(r)-p(r+dr)]$, which repels the mass element $dM$ from the
star center, is balanced by gravitational force $F_g=GdM{\mathcal
M}/r^2$, which pulls the mass element toward the star's center.
Here ${\mathcal M}=4\pi\int_0^rr'^2dr'mn(r')$ is the mass of the
part of the star with radius $r$, and in equations
(\ref{s1}-\ref{s2}) the mass density $mn(r)$ is replaced by energy
density $\varepsilon(r)/c^2$.

The optical potential adds an additional compression force
$F_o=-dV_{o}(r)/dr$ to every neutron in the shell. So, Eq.
(\ref{s1}) should be replaced by another one
\begin{equation}\label{s4}
\frac{dp(r)}{dr}=-G\frac{\varepsilon(r){\mathcal
M}(r)}{c^2r^2}-n\frac{\hbar^2}{2m} 4\pi
\frac{d[nb(n)]}{dr}=-G\frac{\varepsilon(r){\mathcal
M}(r)}{c^2r^2}+\frac{4\pi}{5}(mn_cR_0)^2Gx^3
\frac{d[x^3\beta(x)]}{dr},
\end{equation}
where $R_0^2=5\hbar^2 b(0)/2Gm^3$, and $\beta(n)=|b(E)/b(0)|$.

The pressure $p(r)$ is related to energy density $\varepsilon(r)$
by thermodynamic relationship
\begin{equation}\label{7}
p=-\frac{\partial U}{\partial
V}=n\frac{d\varepsilon}{dn}-\varepsilon=\varepsilon_{c}\varphi(x),
\end{equation}
where
\begin{equation}\label{9}
\varphi(x)=\frac x3\frac{df}{dx}-f=\frac18\left[(2x^2-3)x
\sqrt{1+x^2}+3\ln(x+\sqrt{1+x^2})\right]= x^4\cases{x/5&for
$x\to0$\cr1/4&for $x\to\infty$\cr}.
\end{equation}
It is useful to note that
\begin{equation}\label{9a}
\varphi'(x)\equiv\frac{d\varphi(x)}{dx}=\frac{x^4}{\sqrt{1+x^2}}.
\end{equation}

Let's introduce a unit of space $r_0$, and a unit of mass
\begin{equation}\label{10a}
{\mathcal M}_0=\frac{4\pi}{3}r_0^3\varepsilon_{c}/c^2,
\end{equation}
which will be determined soon, then one can use dimensionless
variables $z=r/r_0$, and $\mu(z)={\mathcal M}/{\mathcal M}_0$.

After substitution of (\ref{6}) and (\ref{7}) into (\ref{s1}) and
(\ref{s2}), one obtains
\begin{equation}\label{s5}
\frac{dx}{dz}\left[\phi'(x)-\frac{4\pi Gmn_c}{5c^2}
R_0^2x^3\frac{d[x^3\beta(x)]}{dx}\right]=-\frac
G{r_0c^2}\mathcal{M}_0f(x) \frac{\mu(x)}{z^2},
\end{equation}
\begin{equation}\label{10xa}
\frac{d\mathcal{M}}{dz}=\mathcal{M}_0\frac{d\mu}{dz}= 4\pi
r_0^3\frac{\varepsilon_{c}}{c^2}z^2f(x).
\end{equation}
Now one can define unit radius $r_0$ by requiring that
\begin{equation}\label{13}
\frac{G\mathcal{M}_0}{r_0c^2}\equiv\frac{4\pi
G\varepsilon_{c}r^2_0}{3c^4}=\frac{4\pi}{3}
n_{c}r^2_0\frac{Gm}{c^2}=1.
\end{equation}
Therefore
\begin{equation}\label{20a}
r_0 = \sqrt{\frac{3c^2}{4\pi Gmn_c}}=
\frac32\frac{\hbar}{mc}\sqrt{\frac{\pi\hbar c}{Gm^2}}= 7.3{\rm\
km}.
\end{equation}
Substitution of (\ref{20a}) into (\ref{10a}) gives
\begin{equation}\label{20b}
{\mathcal M}_0=\frac{4\pi}{3}r_0^3n_{c}m=10\cdot10^{33}{\rm\
g}=5M_\odot.
\end{equation}

Finally Eq-s (\ref{s5}), (\ref{10xa}) can be rewritten as:
\begin{equation}\label{s6}
\frac{dx}{dz}\left[\phi'(x)-\frac{3}{5}\left(\frac{R_0}{r_0}\right)^2x^3
\frac{d[x^3\beta(x)]}{dx}\right]=-f(x)\frac{\mu(x)}{z^2},
\end{equation}
\begin{equation}\label{10xb}
\frac{d\mu}{dz}= 3z^2f(x).
\end{equation}
Substitution of
$$\frac{d\phi(x)}{dx}=\frac{x^4}{\sqrt{1+x^2}},\qquad
\beta(x)=\frac{1}{1+Qx^2}$$ into (\ref{s6}), (\ref{10xb})
transforms them to
$$z^2\frac{dx}{dz}\left(1-\alpha x\frac{\sqrt{1+x^2}(1+Qx^2/3)}{(1+Qx^2)^2}\right)=
-\frac{\sqrt{1+x^2}}{x^4}f(x)\mu(x), \qquad
\frac{d\mu}{dz}=3z^2f(x),$$ where $\alpha\approx13.5$. Integration
of two equations for given $x(0)$ and $\mu(0)=0$ at the star's
center can be easily performed with the help of any existing
software program. The results are presented in table 1.

The first column of the table shows $x(0)$. We made calculations
for five points in the interval $0.1\le x(0)\le0.5$, i.e. for
densities $n(0)=x(0)^3n_c$ in the interval $3.7\cdot10^{36}\le
n(0)\le4.6\cdot10^{38}$ cm$^{-3}$, because for larger densities
the effect of optical potential is negligible, and at smaller
densities the neutron star should contain electron-nuclei plasma.

The table is divided into two parts, the left one contains the
results calculated without optical potential, and the right part
--- with optical potential included. Every part is subdivided
again into two subparts. The left one is calculated with TOV
equations in nonrelativistic Newtonian form, while the right part
is calculated with the system of equations, Eq. (\ref{s1}),
containing general relativity corrections~\cite{re} in the right
hand side:
\begin{equation}\label{s7}
\frac{dp}{dr}=-G\frac{\rho(r){\mathcal M}(r)}{r^2}=
-G\frac{\varepsilon(r){\mathcal M}(r)}{c^2r^2}
\left[1+\frac{p(r)}{\varepsilon(r)}\right] \left[1+\frac{4\pi
r^3p(r)}{{\mathcal M}(r)c^2}\right] \left[1-\frac{2G{\mathcal
M}(r)}{c^2r}\right]^{-1}.
\end{equation}
In every subpart the table contains two columns. The first one
gives dimensionless radius of the star $z_0=R/r_0$, at which
$n(z_0)=0$, and the second column presents dimensionless mass of
the star $\mu(z_0)={\mathcal M}(R){\mathcal M}_0$. Dependence of
$n(r)$ is qualitatively the same as shown in paper~\cite{re}, so
we do not reproduce it here.

\begin{table}
\label{t}
\begin{tabular}{|c|c|c|c|c||c|c|c|c|} \hline\hline
&\multicolumn{4}{|c||}{\bf without
$V_{o}$}&\multicolumn{4}{|c|}{\bf with $V_{o}$}\cr \cline{2-9}
x(0)&\multicolumn{2}{|c|}{\bf
nonrelativistic}&\multicolumn{2}{|c||}{\bf relativistic}&
\multicolumn{2}{|c|}{\bf nonrelativistic}&\multicolumn{2}{|c|}{\bf
relativistic}\cr\cline{2-9} &$z_0$ &$\mu(z_0)$&$z_0$
&$\mu(z_0)$&$z_0$ &$\mu(z_0)$&$z_0$&$\mu(z_0)$\cr \hline\hline
0.1&4.72&0.017&4.70&0.017&4.04&0.011&4.01&0.011\cr\hline
 0.2&3.35&0.048&3.29&0.045&2.96&0.038&2.89&0.036\cr\hline
0.3&2.69&0.085&2.60&0.075&2.45&0.073&2.38&0.067\cr\hline
 0.4&2.31&0.124&2.20&0.101&2.18&0.11&2.1&0.096\cr\hline
0.5&2.04&0.162&1.87&0.121&1.91&0.144&1.79&0.120\cr \hline\hline
\end{tabular}
\caption{ Parameter $x(0)$ in the first column gives density at
the star's center $n=n_cx(0)^3$, parameters $z_0$ give radius of
the star, $R=r_0z_0$, and parameters $\mu(z_0)$ give mass of the
star, $M={\mathcal M}_0\mu(z_0)$. Calculations were made with and
without optical potential, using Eq-s (\ref{s1}) and (\ref{s2}) in
nonrelativistic approximation and using Eq-s (\ref{s7}),
(\ref{s2}) with relativistic corrections.}
\end{table}
From the Table 1 it follows that neutrostriction forces give
corrections to the star's mass and radius, which surpass
relativistic ones.

\section{Neutron star with short range nuclear forces}

At high densities the short range nuclear interactions come into
play. According to (69) of~\cite{re} the energy density of
symmetrical nuclear matter with equal number of neutrons and
protons can be represented as
\begin{equation}\label{n0}
\varepsilon_{sym}=mc^2n_c\left(x^3+0.3x^5-1.5x^6+17x^{9.336}\right).
\end{equation}

The neutron star considered here is not a symmetrical nuclear
matter, because it does not contain protons, and for asymmetrical
nuclear matter energy density according to (86) -- (88)
of~\cite{re} is
\begin{equation}\label{n00a}
\varepsilon_{nonsym}=\varepsilon_{sym}+\Delta\varepsilon,
\end{equation}
where $\Delta\varepsilon$ can be represented as
\begin{equation}\label{n0a}
\Delta\varepsilon=mc^2n_c(0.07x^5+0.55x^6).
\end{equation}
Therefore the total nuclear energy density for neutron matter is
\begin{equation}\label{n00b}
\varepsilon_{nonsym}=\varepsilon_{sym}+\Delta\varepsilon=
mc^2n_c\left(x^3+0.37x^5-0.95x^6+17x^{9.336}\right).
\end{equation}
Now we want to compare the attractive part of nuclear energy
density, $\varepsilon_{-}=-mc^2n_c0.95x^6$ with optical energy
density $\varepsilon_o=mc^2n_c4.6x^6/(1+Qx^2)$. The ratio of
optical energy density  to the attractive part of nuclear energy
density is
\begin{equation}\label{n3}
\frac{\varepsilon_{opt}}{\varepsilon_{-}}\approx\frac{5}{1+Qx^2},
\end{equation}
and we see that optical energy is larger than the nuclear one at
$Qx^2<4$, or $x<0.12$, which is equivalent to
$n<0.0017n_c=0.04n_0$.

\section{Discussion}

The neutrostriction forces are to be taken into account in
calculation of neutron star. They play important role at low
densities and small masses ${\mathcal M}<M_\odot$. More over they
present many interesting problems worth of research for their own.

\begin{enumerate}
\item It seems that neutron-neutron interaction in the degenerated
neutron gas is eliminated because of the Pauli exclusion
principle. However the Pauli exclusion principle, as correctly is
pointed out in~\cite{bom}, eliminates scattering process and
imaginary part of the scattering amplitude, but it does not affect
its real part. It means that because of the Pauli exclusion
principle the optical potential in neutron stars becomes lossless.
\item The decrease of $b(n)$ with density is a
source of pulsations, and it is interesting to investigate how
well possible pulsation match parameters of the observed pulsars.
\item The pulsations are especially well understandable,
if scattering contains a resonance, as is shown in fig. \ref{ff3}.
At some energy the scattering amplitude changes the sign, so for
smaller density the optical potential is attractive, and for
larger one it becomes repulsive. At the point $E$, where $b(E)=0$,
pulsations arise naturally.
  \begin{figure}[ht]
{\par\centering\resizebox*{10cm}{!}{\includegraphics{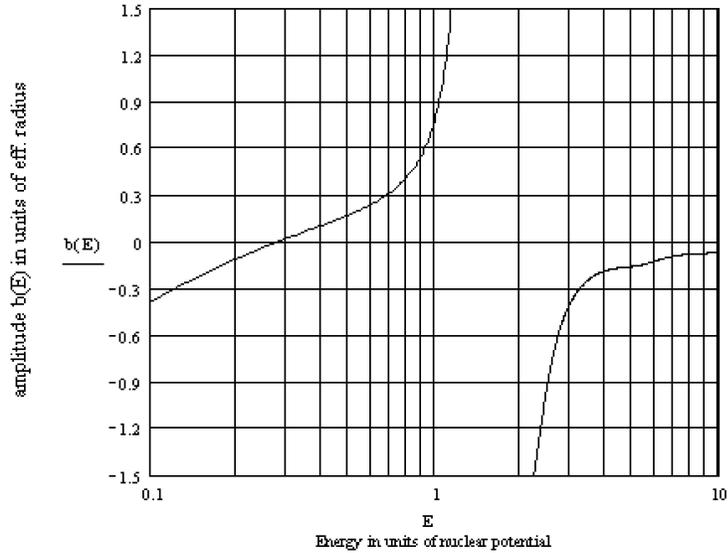}}\par}
\caption{\label{ff3}Schematic energy dependence of scattering
length in presence of resonance.}
\end{figure}
\item In the case of resonance we have a mechanism for star explosion.
Indeed, if at contraction the density (and therefore Fermi energy)
overcomes the resonant point, the strong repulsive optical energy
changes to strong attractive one, and a huge energy is released.
\item Protons were not considered in this evaluation,
however inclusion of protons will not change the arguments about
optical potential, because the coherent neutron-proton scattering
amplitude is nearly the same as the neutron-neutron one. At the
same time, with protons one must also take into account
neutron-proton resonances, which do certainly exist. Their
presence can also provide a source of pulsations and explosions.
\item We considered only s-scattering amplitude. However,
when the energy (or density) increases, one must also include p-,
d- and higher harmonics. The question arises: how will they affect
the results.
\item This paper discussed only neutron stars,
but the notion of the optical potential is considerably more
widely applicable. It can be used in other stars, in Bose-Einstein
condensates, superfluidity and superconductivity, because
everywhere we must take into account the coherent atom-atom and
atom-electron scattering amplitude.
\end{enumerate}

\section*{Acknowledgment}
The author is grateful to Yu. Petrov, V.L. Lyuboshits and Prof.
R.R. Silbar for interesting discussion, to B.V. Vasil'ev and S.B.
Borzakov for their useful remarks, to E.Shabalin for his support,
to Steve Lamoreaux for collaboration, to I. Carron and I.Petruski
for their help.

\section{Hystory of submissions and rejections}

I submitted the paper first to PRL, and did not fight against
their rejection. I did not save the referee report, however his
comment was like the following one: ``I am not a specialist in
neutron star physics, but I do not believe that such simple things
are not known to them, so I recommend to reject the paper.

On 04.05.2005 after having read their paper~\cite{re} I submitted
the paper to Amer. J. Phys. I had 3 referees and fought hard.

0n 03.08 The 1-st referee conclusion was: ``Because the author
wishes to introduce a new idea and several new conclusions, the
paper (after extensive rewriting into better English and, of
course, and with less didacticism) might be appropriate for
another journal.''

I corrected didacticism and English (with all my best), and
pointed out that the idea about neutrostriction is published in my
book~\cite{ign}, however on 19.08 the referee conclusion was the
same:

``Although Dr. Ignatovich is highly respected for his
accomplishments, this paper is still misdirected...''

I insisted for further refereeing and on 26.09 received two
reports

REVIEW \#1:

``I agree with the reviewer that this paper constitutes an
argument for "new" physics. The assertion of this paper is that
"neutrostriction" results in an additional term in the standard
Tolman-Oppenheimer-Volkov equation. I can't find anybody other
than the author who uses the term neutrostriction, and it seems a
pretty big claim that the astrophysical community has been missing
this potential for so many years. The author is well-published in
the field of neutron scattering, and may be correct that the
astrophysical community needs to take account of coherent neutron
scattering effects, but AJP isn't the place to be putting forward
such claims.''

REVIEW \#2:

''My main objection continues to be the misplacement of this
submission.  The very first line of the paper's abstract states,
"It is demonstrated that...".  This is the announcement of a new
result by the author (even if he has already made the announcement
before in any number of conferences and published conference
papers).

Following the AJP guidelines which state, in part, "Manuscripts
announcing new theoretical or experimental results ... are not
acceptable and should be submitted to an archival research journal
for evaluation by specialists," I must again recommend that the
AJP not publish this paper.''

I appeal directly to Jan Tobochnik, the Chief Editor of AJP,
however in vain. The last report was on 20.10:

``I completely agree with the previous referee regarding
unsuitability of this paper for American Journal of Physics. What
is presented here looks like a new result and it needs to be
reviewed by a technical expert for an archival journal such as
Physical Review C. I recommend that the article is rejected
without prejudice, i.e. the subject matter is more suitable for a
technical journal in the field in nuclear physics. I did not find
the article particularly pedagogical.''

So on 26.10 I submitted the paper to Phys. Rev. C

\subsection{Referee report on 10.11.2005 and my replies in italic}
 Dear Dr. Ignatovich,

The above manuscript has been reviewed by one of our referees.
Comments from the report are enclosed.

These comments suggest that the present manuscript is not suitable
for publication in the Physical Review.

Yours sincerely,

Jonathan T. Lenaghan Assistant Editor Physical Review C

----------------------------------------------------------------------

Report of the Referee -- CKJ1005/Ignatovich

----------------------------------------------------------------------

Everything in this paper is just wrong and it should never be
published. Bad language set aside,

{\it I edited it once again and will be grateful for any
suggestion.}

the physics presented here is a collection of bits and pieces from
here and there, with the attempt to invent a name,
neutrostriction, for a force deriving from well-known many-body
physics, that only the author seems to use. The paper carefully
avoids referring to the literature in many occasions. A search on
google.com for neutrostriction shows that this is a term used only
by this author.

The discussion of section 4 sets the standard. It suffices to read
point 1, page 10. Here the author states

'in some textbooks it is claimed that the neutron-neutron
interaction in the degenerated neutron gas is absent because of
pauli exclusion principle'

There is a careful omission to any reference for obvious reasons.

{\it I included reference to Bohr and Mottelson, and changed the
paragraph.}

There are, to my knowledge, no textbooks which claim such things.
It is the pressure set up by a gas of identical particles (here
neutrons) interacting via, at least, the neutron-neutron
interaction. This interaction leads, due to the Pauli principle to
a repulsive EoS which counteracts gravity. Read carefully Shapiro
and Teukolsky, The physics of compact objects, chapter 8 and 9.
This is basic quantum mechanical wisdom.

The author continues 'it means that scattering cross section,
which enters imaginary part of the scattering amplitude is
suppressed, however, the real part is no changed, so pauli
exclusion principle helps to create lossless optical potential'

The real part of the optical potential is related, via a
dispersion relation to the imaginary part. Any quenching of the
imaginary part leads to changes in the real part. The statement
above is simply wrong.

{\it The referee speaks about relation between imaginary and real
parts of complex analytical functions. I speak about a complex
number, which can have arbitrary real and imaginary parts. The
potential contains scattering amplitude with imaginary part
defined by the optical theorem. In some respect the real  part
also changes with the change of imaginary one, but this change can
be neglected.}

The optical potential is an outcome of the many-body physics (see
standard texts such as he many-body book of Fetter and Walecka),
in this case the interactions among neutrons in an idealized star
of neutrons only. Viz, it is derived from many-body physics.

{\it No. Many-body theories do not include multiple wave
scattering, which transforms short-range interaction into a
long-range one.}

The author, both in the abstract and the introduction, portrays
the optical potential as something independent from the many-body
physics and even invents a name for it, neutrostriction.

There is no new mechanism or physics at play here, the optical
potential is entirely linked with the many-body physics, which in
turn sets up a repulsive EoS (due to the Pauli principle) which
counteracts gravity.

{\it No! Pauli exclusion principle and repulsive forces are
included in my consideration, however, there is also an attractive
long range interaction, which is overlooked by many-body
theories.}

The rest of the discussion in this paper is more or less at the
same level and I refrain from further comments.

{\it Dear editors, consider, please, my objections against main
referee points, and accept, please, my paper for publication.}

\subsection{Referee report on 30.01.2006 and my replies in italic}

The same formal rejection letter but from

Christopher Wesselborg Associate Editor Physical Review C

----------------------------------------------------------------------

Second Report of the Referee -- CKJ1005/Ignatovich

----------------------------------------------------------------------

I thank the author for his reply. I still disagree with the
author, especially his answers to remarks 3 and 4 below.

3. ``The optical potential is an outcome of the many-body physics
(see standard texts such as he many-body book of Fetter and
Walecka), in this case the interactions among neutrons in an
idealized star of neutrons only. Viz, it is derived from many-body
physics.'

No. Many-body theories do not include multiple wave scattering,
which transforms short-range interaction into a long-range one.

-- I disagree strongly with this statement; if performed
correctly, including particularly particle-hole correlations one
generates long-range correlations. Many-body theory such as
Green's function Monte Carlo, coupled cluster theory or the
summation of parquet diagrams, should include these correlations.

The optical potential can in turn be derived from many-body
theories, see, e.g., Fetter and Walecka, Quantum theory of
many-particle systems, chapter 40, pages 352-357.

{\it Can be derived or is derived? Pages 352-357 do not help. Do
you know, that though neutron-nuclus potential $V(z)$ is negative,
the optical potential of the medium composed of nuclei has
positive, i.e. repulsive optical potential? In particular, all the
terms on pages 352-357 are proportional to negative value $V(z)$
and cannot become positive.}

4.``There is no new mechanism or physics at play here, the optical
potential is entirely linked with the many-body physics, which in
turn sets up a repulsive EoS (due to the Pauli principle) which
counteracts gravity.'

{\it No! Pauli exclusion principle and repulsive forces are
included in my consideration, however, there is also an attractive
long range interaction, which is overlooked by many-body
theories.}

-- See my reply to the remark 3. I don't see why this shouldn't
arise from a many-body description as offered by, for example,
Green's function Monte Carlo or the other methods mentioned above.
I cannot see any new physical mechanism at play.

{\it See my reply above at remark 3, which proves that many-body
theory does not describe multiple wave scattering phenomenon,
which accounts for neutrostriction.}

I would like to uphold most of my previous criticism. The
disagreements voiced in the previous report remain and I would
advise the author to ask for a new referee if he feels incorrectly
judged by me.

{\it Dear editor I don't agree with judgement of the referee, I
ask you to reconsider your decision and to publish my paper, which
is absolutely correct and discovers new phenomena.}

\subsection{Referee report on 10.02.2006 and my replies in italic}

{\it Dear editors, according to rules of Physical Review I have
right to appeal. I replied to all referees, and pointed out what
do they not understand. No referee could raise an objection
against my arguments, they only don't believe that it is possible
to say a new word after their 50 years research. It is wrong. I
insist that my paper is absolutely correct and discovers new
phenomena. Consider, please, this letter as my appeal.}

 The above manuscript has been reviewed by one of our
referees. Comments from the report are enclosed.

We regret that in view of these comments we cannot accept the
paper for publication in the Physical Review.

In accordance with our standard practice, this concludes our
review of your manuscript.  No further revisions of the manuscript
can be considered.

Yours sincerely,

Benjamin F. Gibson Editor Physical Review C Email:
prc@ridge.aps.org Fax: 631-591-4141 http://prc.aps.org/

----------------------------------------------------------------------

Report of the Second Referee -- CKJ1005/Ignatovich

----------------------------------------------------------------------

I agree with the previous referee in all respects, so I have
nothing to add that would be of interest to the editors.

But perhaps if I restate a piece of the referee's comments in my
own words it will be helpful to the author in understanding our
point: At its most primitive level, say the state of the art of
half a century ago, the calculation of the nuclear interaction
energy would have been the Hartree term, which is no more or less
than the expectation value of the N-N potential in the Fermi gas,
leaving out the exchange terms (which are always a lot  smaller).
An optical potential, or an index of refraction  derived from the
Born forward scattering amplitude, generates exactly the same
term.

{\it The Born terms in strong interactions are unable to give a
correct value and even sign for a scattering amplitude. This
scattering amplitude gives you not a nuclear optical potential,
but optical potential of the full medium of nuclei.}

And the same is true in principal if we use data to directly make
the estimate of the nuclear interaction energy, instead of using
data to find the potential as an intermediate step.

Thus whatever physics is there in the author's optical potential
is contained in the work of the people who have calculated the
equations of state used for neutron stars. These workers have used
the present state-of-the-art  nuclear potentials, fit to vastly
more data that the author  invoked, and applied in many-body
calculations that have  grown more sophisticated over the years.
We certainly hope,  or even trust, that there is much, much more
correct physics  in these equation of state results in the
literature than in the  kinds of estimates one could make 50 years
ago.

{\it I present a good physical idea which shall enrich equation of
state and sophisticated state-of-the-art  nuclear potentials.}

\section*{Correspondence about the
appellation}

\subsection*{Letter on 03.03.06 from Dr. Christopher Wesselborg}

Dear Dr. Ignatovich:

Thank you for your prompt response. I suggest that you send us
your appeal letter with your response to the referees' criticisms,
part of which seems to be already contained in your  recent email.

As to your specific inquiry, I was referring solely to your
previous, rather general (i.e., unspecific) request for an appeal.
Note that the editors rejected your manuscript based on the
reports from the two referees and your responses.

For your information, we had also sent the previous correspondence
to the second referee, including your resubmission letters, when
we asked a second referee for an additional opinion.

Again, we will begin with the appeal as soon as we have your
complete appeal letter.

Sincerely,

Christopher Wesselborg

Associate Editor

Physical Review C

\subsection*{My reply on 06.03}
Dear Dr. Christopher Wesselborg, I need your advice, how to make
The Complete appeal letter? I feel that everything depends whether
it will be correct or not. Thank you for your information that the
second referee had my replies to the first one. I wonder why he
did not take them into account.

How to make the complete appeal letter? Should I analyze arguments
of both referees? Should I add more arguments in defense of my
position? There is a single point of our disagreement: They insist
that everything is contained in many-body theory. The first
referee even pointed out the pages of the many-body book, and the
second referee told that everything is contained in Born forward
scattering amplitude. My point is: that the values on the pages,
the first referee pointed out, and the Born scattering amplitude
of the second referee are proportional to the two-body interaction
potential. Therefore they have the same sign as the interaction
itself. The multiple wave theory contains not the Born scattering
amplitude. The scattering amplitude is the result of more rigorous
solution of scattering problem for a given two-body potential.
This amplitude can be of opposite sign than the potential. The
optical potential of medium is a secondary construction, which
uses multiple-wave scattering formalism, absent in many-body
theory.

The second referee does not accept my paper also because my idea
is very simple comparing to sophisticated theories used by present
day astrophysicists. I remind you that some referees in other
journals rejected the paper, because they did not believe that
such simple things are not known to astrophysicists. You see, they
are really not known!

Dear Dr. Christopher Wesselborg, may I ask you, are the above
arguments appropriate for the Complete appeal letter? Should I
write the similar letter via internet resubmission? What is the
form of such a Complete appeal letter? Really yours, really need
your help, Vladimir Ignatovich.

\subsection*{Reply on 08.03}

Dear Dr. Ignatovich,

Thank you for your message of March 6.  I appreciate that you want
to write the most complete appeal letter possible.  You should
analyze and carefully consider the arguments of the referees.  In
particular, you should try to address in your letter each point
made by both referees. Please avoid polemical language and argue
your case dispassionately. If you have more arguments in defense
of your position then you should make them.  Note, however, the
appeals process is based on the rejected version of the manuscript
and the further revisions are not considered.

You should submit your appeal letter via the internet submission
server if possible.  If I can be of any more help, please let me
know.

Yours sincerely,

Jonathan T. Lenaghan

Assistant Editor

Physical Review C

\section*{Appellation 16.03}

The main objection of two referees, formulated in my own words, is
the following: ``the optical potential, which leads to refraction
index, can be found in many-body theory. This sophisticated theory
does not see the effects discussed in your paper, therefore your
ideas are wrong.``

My defence was: ``the many-body theory does not contain multiple
wave scattering phenomenon, because in other case it would found
the potential I discuss in my paper.''

Now I must admit that I was not right. Thanks to the first
referee, who pointed out to me the book by Fetter and Walecka
[FW], I could improve my education. Now I can tell that many-body
theory contains everything I discuss in my paper, nevertheless the
effect was overlooked by astrophysicists and I can explain why.

First I want to point out the place in FW, where this potential is
shown. It is formula (11.65) for chemical potential $\mu$,
obtained by V.M.Galitskii:
$$\mu=\frac{\hbar^2k_F^2}{2m}\left[1+\frac{4}{3\pi}k_Fa+\frac{4}{15\pi^2}(11-2\ln2)(k_Fa)^2\right],
\eqno(11.65)$$ where $a$ is scattering length, and $k_F$ is Fermi
wave-number: $k_F^3/3\pi^2=n$ is particle density.

This formula was obtained for dilute Fermi-gas, when $k_Fa\ll1$,
which case is just what I discuss in my paper. If we neglect last
term $\propto (k_Fa)^2$, we can rewrite (11.65) in the form
$$\mu=\frac{\hbar^2}{2m}k_F^2+\frac{\hbar^2}{2m}4\pi na,\eqno(I)$$
and the second term is just optical potential which I introduced
in (1) (in my notation scattering length $a$ is $b$).

However formula (11.65) was found for scattering from a repulsive
core, when the actual (not perturbative) scattering length $a$ is
positive. So the optical potential $\propto4\pi na$ is also
repulsive.

Attractive, negative, potentials are not considered by many-body
theory because, according to problem 1.2 of the chapter 1, a
system with a potential $V(\rr)<0$, and $\int|V(\rr)|d^3r<\infty$,
will always collapse. The collapse follows from expression (I),
because for negative $a$ and high density $n$ chemical potential
becomes $\mu\approx Cn^{2/3}-an$, where $C$ is a constant, which
goes to $-\infty$ when $n\to\infty$.

It is correct for constant $a$, but it is not correct, if we take
into account energy dependence of $a$. My formulas (7) and (8)
introduce dependence $a/(1+k_F^2|a|r_0)$, where $r_0$ is effective
radius of interaction, so we have no collapse. More over, if
neutron-neutron interaction contains a repulsive core, the
scattering length can become positive at high density.

I can summarize as follows: The sophisticated mathematics contains
everything, but without physical idea it is difficult to predict
something. On the other hand, a physical idea helps to predict
with few relevant mathematics, however, and it is especially
important, the correct idea is always supported by sophisticated
mathematics. My paper contains idea, it helps to predict some
phenomena and it is supported by many-body theory. I think it is
an important contribution both: to physics of neutron stars and to
many-body theory.

\subsection*{Reply to the appellation 12.04}

Dear Dr. Ignatovich,

This is in reference to your appeal on the above mentioned paper.
We enclose the report of our Editorial Board member Richard
Furnstahl which sustains the decision to reject.

Under the revised Editorial Policies of the Physical Review (copy
enclosed), this completes the scientific review of your paper.

Yours sincerely,

Benjamin F. Gibson Editor Physical Review C Email:
prc@ridge.aps.org Fax: 631-591-4141 http://prc.aps.org/

----------------------------------------------------------------------

Report of the Editorial Board Member -- CKJ1005/Ignatovich

----------------------------------------------------------------------

I concur with the comments of the first and second referee.  The
physics discussed in this manuscript is not new and is presented
in a misleading way (e.g., the comparison of gravitational and
"optical" energies using the scattering length only throughout the
volume of the neutron star).

Based on the reports of the referees and my own assessment, I
recommend that this manuscript should not be published in Physical
Review C.

  Richard Furnstahl

  Memeber, Physical Review C Editorial Board

Please see the following forms:

  http://forms.aps.org/author/polprocc.pdf
    PRC EDITORIAL POLICIES AND PRACTICES

\subsection*{My reply 13.04}

Dear Editor!

No argument is an argument for Dr. Furnstahl, who "concurs with
the comments of the first and second referee" without an argument.
He writes that "The physics discussed in this manuscript is not
new". Then how does he concur with the statement of the first
referee: "Everything in this paper is just wrong"? He writes that
my not new physics "is presented in a misleading way (e.g., the
comparison of gravitational and "optical" energies using the
scattering length only throughout the volume of the neutron
star)." I cannot understand neither this sentence, nor what is
misleading in such a comparison? May I ask you: do you understand?
May I ask you to explain it to me?

Now, when everything is in vain, may I ask you to send all my
files to Editor-in-Chief Dr. Blume? I know that his reply will be
negative and formal. I can even formulate his reply, but will not
do that. Let his secretary to use a template to support the
decision of the editorial board and to blame me for insulting
manners. His reply will not help, but I need it as a last stone
for a monument to American Physical Society.

Vladimir Ignatovich

\subsection*{From Phys.Rev. 17.05}

Dear Dr. Ignatovich,

Thank you for your message.  We will soon initiate the appeal of
your manuscript to the Editor-in-Chief.  We are writing to ask you
to draft an appropriate cover letter.  Your current letter may be
interpreted as polemical. Your appeal letter should clearly
demonstrate why your manuscript warrants publication in view of
the arguments presented by the referees, Editors and the Editorial
Board member.

Upon receiving your cover letter, we will initiate your appeal to
the Editor-in-Chief.  Thank you for your attention to this matter.

Yours sincerely,

Jonathan T. Lenaghan Assistant Editor
\subsection*{My reply on 18.05}

Dear Dr. Lenaghan, I was really surprised to get your friendly
letter after so long silence. May I ask you to teach me, how to
compose such a letter. I prepared it, but I am not sure it has an
appropriate form. I has a terrible experience that no appeal is
successful after rejection by editorial board. Nevertheless, I am
ready to try and try again. May I ask you to help me? Read,
please, my reply, and give me to know, please, what is it better
to change. Yours sincerely Vladinir.

To Dr. Blume.

Dear, Dr. Blume, I appeal to you as the Editor-in-Chief of the
American Physical Society, against rejection of my paper titled:
"Neutrostriction in neutron stars" by editorial board of Phys.
Rev. C.

In this paper I had shown that neutron-neutron scattering forms
strong attractive optical potential inside neutron stars, which
compresses the star together additionally to gravity. This
attraction decreases with increase of density, but leads to many
interesting physical effects and influences  such parameters of
neutron stars, as radius, mass and distribution of density. Effect
of this optical potential can surpass effects of general
relativity.

My paper was considered by two anonymous referees and by a Memeber
of Physical Review C Editorial Board Richard Furnstahl. So
formally my paper met a fair hearing. However the reports of all
these referees show that they did not consider my paper
responsibly. May I ask you to look, please, at their arguments and
my responses.

The first report of the first referee started with the words:
"Everything in this paper is just wrong and it should never be
published." He pointed out several "errors" and claimed that
"There is no new mechanism or physics at play here, the optical
potential is entirely linked with the many-body physics, which in
turn sets up a repulsive EoS (due to the Pauli principle) which
counteracts gravity." The last sentence clearly shows presumption
of the first referee. I calmly replied. Included some references,
which the first referee supposed to be omitted intentionally, and
explained to the referee all his misunderstandings with respect to
"errors". His second report was softer, however he insisted on his
presumption. Our difference was: Referee claimed that many body
contains everything, and it does not show my effects, while I
insisted that many body theore is incomplete, because it does not
show my effects.

The second referee also rejected my paper on the same presumption.
His report started with the words: "I agree with the previous
referee in all respects, so I have nothing to add that would be of
interest to the editors." He tried to teach me that the state of
the art of calculations became more sophisticated than my approach
which is alike to the old-fashoned Hartree approach to the many
body theory. He wrote that optical potential and an index of
refraction can be found from the Born forward scattering
amplitude. The truth of this reply is not complete. The scattering
amplitude can be derived from precise equations and it can differ
in sign from the Born amplitude.  Such a difference is crucial for
determination of the sign of the optical potential.

I used my right to appeal, and during preparation of my appeal
letter, I studied more carefully the many-body theory and found
that in principle it really contains the attractive optical
potential, but it was overlooked by scientists, and I even
understood why. I pointed out it in my appeal letter. I found also
that the error in many body theory is related to the widely
accepted practice of discretesation. We all introduce finite
dimension L, when we describe scattering. It seems very natural,
but I found the first case where this practice leads to an error.
I pointed it out.

In reply to my appeal letter, the member of Editorial board Dr.
Richard Furnstahl did not discuss the point I mentioned. His
report starts with the words: "I concur with the comments of the
first and second referee.  The physics discussed in this
manuscript is not new and is presented in a misleading way." These
words are not understandable. If it is not new physics, why the
neutrostriction forces are not discussed by astrophysical
community? If it is not new physics, why the first referee, with
whose report Dr. Furnstahl concurs, claimed that it is wrong
physics?

Dear Dr. Blume, I would be very grateful to you if you find a time
or ask some other experts to explain me what did Dr Furnstahl
meant. I still continue to think that my paper is a very important
contribution to the neutron star physics and to many-body theory.

Yours sincerely, Vladimir Ignatovich.

\subsection*{From Phys.Rev. on 06.06}
Dear Dr. Ignatovich,

Thank you for your improved cover letter.  I will be forwarding
your file to the Editor-in-Chief very shortly.  If I can be of any
more assistance, please feel free to let me know.

Yours sincerely,

Jonathan T. Lenaghan Assistant Editor

\subsection*{Letter from Martin Blume, dated 09.06}

Dear Dr. Ignatovich,

I have reviewed the file concerning this manuscript which was
submitted to Phys. Rev. C. The scientific review of your paper is
the responsibility of the editor of Phys. Rev. C, and resulted in
the decision to reject your paper. The Editor-In-Chief must assure
that the procedures of our journals have been followed responsibly
and fairly in arriving at this decision.

Contrary to your assumption, every appeal case that is submitted
to me receives a thorough review. I note that the referee and
editorial board member were unanimous in their opinion that your
paper was not appropriate for Phys. Rev. C. Let me add that we
take pride in our appeal process. Many other journals have no such
policy; for those journals an editorial rejection is final and
authors have no right to appeal.

On considering all aspects of this file I have concluded that our
procedures have in fact been appropriately followed and that your
paper received a fair review. Accordingly, I must uphold the
decision of the Editors.

Yours sincerely, Martin Blume.

\subsection*{My reply to it on 27 June}

Dear Dr. Blume,

Thank you for your reply and for taking time to review my case. I
continue to be strongly convinced, that my paper contains
important work, which can and should be allowed to appear in Phys.
Rev. C. I still believe that the referees unfortunately were not
well qualified to review my manuscript, because from their
comments I deduce that they did not understand my work.

I have a concern about the appeal process, namely: were there any
precedents when the opinion of editorial board member was opposite
to unanimous opinion of two referees? Were there any precedents
when your decision was opposite to that of the editorial board? If
not, the appeal process seems to fail, as the unanimous opinion of
two referees predefines the appeal outcome.

To overcome this possible flaw, I propose to send my file to an
independent person, who would agree to judge the validity of
arguments of both sides.

I would like to suggest a person, who is to my opinion qualified
to listen and understand the arguments. Furthermore, it will be
even better if you could also choose one, and then compare the
judgments of these two people  to help you to come up with the
final decision. I understand that it will require some of your
time, but it will be well rewarded by the benefit to science.

I will highly appreciate your attention and effort to resolve my
case.

Sincerely, Vladimir Ignatovich

P.S. Please contact me via e-mail ignatovi@nf.jinr.ru


\begin{thebibliography}{9}
\bibitem{ig0}
V.K. Ignatovich, {\it Multiple Wave Scattering Formalism and the
Rigorous Evaluation of Optical Potential for Three Dimensional
Periodic Media.} Proc. of the International Symp. on Advance in
Neutron Optics and Related Research Facilities. (Neutron Optics in
Kumatori '96) J. Phys. Soc. Japan, v. 65, Suppl. A, 1996, p. 7-12.
\bibitem{ign}
V.K.Ignatovich, {\it The physics of ultracold neutrons.} Clarendon
Press, Oxford, 1990.
\bibitem{huhn}
Huhn V V, Watzold L, Weber C, Siepe A, von Witsch W, Witala H, Glockle W.,
New attempt to determine the n-n scattering length with the 2H(n, np)n reaction.
Phys Rev Lett. 2000 Aug 7;85(6):1190-3.
\bibitem{conf} V.K.Ignatovich, in: Neutron Spectroscopy, Nuclear Structure,
Related Topics; ISINN-12 Dubna May 26-29, JINR 2004, pp. 117-132;
Proceedings of the XXXVII-VIII winter school in PINPI on Physics
of atomic nuclei and elementary particles, St.Petersburg, 2004,
pp. 446-466.
\bibitem{re}
Richard R. Silbar \& Sanjay Reddy, Am.J.Phys. v. 72 (7) 892-905
(2004)
\bibitem{bom}
A. Bohr, B.R. Mottelson, Nuclear structure. (Nordita, Copenhagen,
1969)Ch. 2, \S 3, between formulas (2.220) and (2.221).
\end{thebibliography}
\end{document}